# Performance Study of IEEE802.11e QoS in EDCF-Contention-based Static and Dynamic Scenarios

[*]Khaled Dridi, [*]Nadeem Javaid, [*,$]Karim Djouani, [*]Boubaker Daachi

[*]LISSI, University of Paris East, France,
[$]F'SATIE, TUT, South Africa,
khaled.dridi@etu.univ-paris12.fr
nadeem.javaid@etu.univ-paris12.fr

*Abstract*—In this paper, we carry-out a study of the Quality of Service (QoS) mechanism in IEEE802.11e Enhanced Distribution Coordination Function (EDCF) and how it is achieved by providing traffics with different priorities. It can perform the access to the radio channel or just simply it can considerably be declined subsequently to a variation of network dynamicity. The results of the proposed analysis show that the EDCF scheduler looses the ability of the traffic differentiation and becomes insensitive to the QoS priority requirements. Consequently, it goes away from the region of stability and EDCF doesn't offer better performance than the conventional DCF scheme. Therefore, traffic specifications are weakly applied only for the channel occupation time distribution. During the handoff between the Base Stations (BS's), the response time of the data rate application within the roaming process grows to the initial throughput level. Performance metrics at the MAC layer, like throughput, End-2-End delay, and packet loss have been evaluated.

## I. INTRODUCTION

Being flexible, easy deployable, mobile and capable for high transmission data rate, wireless networks are superior to wired ones but they do have some pains, like shared bandwidth, variable delays, high bit-error rates, etc. IEEE 802.11 standard [1] uses a shared medium and has some problems, like, congested medium, uncertain collision rates and traffic differentiation. IEEE 802.11 has a mode of operation which provides service differentiation, but it shows poor link utilization, so many researchers have contributed to solve this problem by proposing several schemes for supporting quality of service. The conventional IEEE802.11 MAC-layer incorporates an optional access method known as Point Coordination Function (PCF), which is used in Contention Free Period (CFP) on infrastructure network configurations. It is based on poll-and-response mechanism. For Ad-hoc configuration, the Distributed Coordination Function (DCF) is the fundamental access method of the IEEE802.11 MAC which applies the Carrier Sense Multiple Accesses with Collision Avoidance (CSMA/CA) technique. This access method has no priority for real time traffics [2]. A single first-in-first-out (FIFO) transmission queue is imposed on 802.11 MAC depending upon the status of the channel which can be occupied by the traffic transmission or it can simply be idle.

DCF with its mandatory CSMA/CA starts working when there are a number of active stations and among them one station sends a frame which already arrived to the head of the transmission queue [3]. Now, there are two possibilities; the first when the channel is busy and the second when it is idle. In earlier case, MAC waits until the medium becomes idle and defers for an extra interval of time, DCF Inter-frame Space (DIFS). If this period of time is enough for keeping the channel idle, the CSMA starts to process. It activates the backoff by selecting a random Backoff Counter (BC) and decrements this BC for each passed time slot, during which the medium stays idle. Finally, when the BC becomes zero, the frame is transmitted. In the same case when the channel is busy and the MAC is in either the DIFS deference or the random backoff process, if a frame arrives at the head of the queue, the above explained process is again applied.

In the situation that the idle medium is not longer than the DIFS time and there is no on-going backoff process, when a frame reaches an empty queue, it is transmitted. A station immediately acknowledges on successful reception after a short-IFS (SIFS). If a frame is received successfully and its acknowledgment is not received, the sender assumes that the frame needs to be retransmitted after another random backoff [4]. After the end of DIFS time, all other stations resume the backoff process.

The rest of the paper is organized as follows: Section II summarizes the IEEE 802.11e: EDCF Contention Based Access and QoS mechanism in wireless LAN's. Section III describes three link metrics. Section IV gives description of the three scenarios designed for the simulation. Section V and VI are devoted for the conclusion and references respectively.

## II. IEEE 802.11e STANDARD SPECIFICATIONS

### II. 1. EDCF CONTENTION BASED ACCESS

Original IEEE 802.11 is not capable of differentiating frames with different priorities rather it provides an equal chance to all stations contending for the channel access for transmission. The access method in the new IEEE 802.11e is called Hybrid Coordination Function (HCF), it combines functions from both the DCF and the PCF [8].

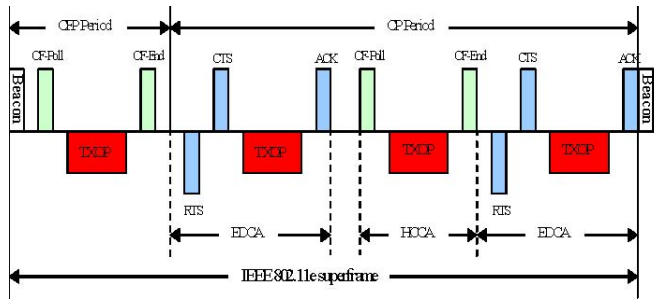

**Fig.1. HCF Super-frame Structure [5]**

To provide a mechanism based on distributed access for service differentiation and to improve the contention-based access mechanisms of IEEE 802.11, Enhanced DCF was proposed. Fig. 1. shows the super-frame of HCF presented in [5]. The great challenge was to make sure that EDCF should be compatible with the old DCF since large number of devices complying with the old standard had been deployed.

The new mechanism classifies the traffic into 8 user classes, with the modified size of Contention Window ($CW_{min}$) and the inter-frame spaces. Smaller the contention window, shorter will be the backoff intervals. Therefore, the traffic priority will be greater. An inter-frame space, Arbitration Inter-frame Space (AIFS) is introduced to start decrementing the backoff timer as in ordinary DCF or to avoid waiting a DIFS before trying the access to the medium. AIFS is associated with each traffic class and is evaluated as a DIFS plus a number of time slots. It implies that traffic using a large AIFS will be assigned lower priority.

To achieve a better utilization of the wireless medium and in order to enhance the performance, IEEE 802.11e may possibly use *packet bursting*. It consists of allowing a station to send more frames once it has gained the access to the idle medium through ordinary contention during *Transmit Opportunity (TXOP)*; an interval of time which is bounded and during which a mobile node can send as much number of frames as it wants (but duration of the transmissions must not exceed the maximum period of the TXOP limit. The packet burst is terminated, if a collision occurs or no acknowledgment frame is received [7]. The most priority traffic operates with Short Inter-frame Space (SIFS), which is the small time interval between data-frame and ack-frame.

### II. 2. QoS in WLAN

QoS is a mechanism which satisfies the service performance across a network or it is an ability of a network to provide desired handling of the traffic which meets the expectations of the end applications. If a network, for example, supports a set of traffic specifications, as, bandwidth, transmission delay, jitter, etc, then it is said to support QoS delivery. It treats packets of all traffic categories at the same priority, that is, no service differentiation at all. At the end, all sorts of traffics, including voice and video traffics, suffer from delays and bandwidth variations.

## III. MAC-Layer-METRICS

The metrics we have used in our evaluation are throughput, end-2-delay, and packet-loss.

### III.1. Packet Loss

The average packet loss for the stations shows that how well the QoS schemes can provide service differentiation between the various priority levels.

### III.2. Throughput

The normalized throughput is calculated as the percentage of the offered data rate that is actually delivered to the destination.

### III.3. End-2-End delay

The reason for studying average end-2-end delay is that many real-time applications are very sensible to high delays, after which the data will be useless. It is, therefore, very important for real-time flows to be provided provide with low delay. Because this class of traffic often has a delay bound after which the data is useless, it does not suffice to just study the average access delay, since the average might be rather low even if a large part of the packets have unacceptable delays. We present the cumulative distribution of the end-2-end delays for all priority of flows if there is not enough traffic to accommodate the polls.

In the graphs of Fig. 2, we present the graphs of End-2-end delay for only the 3rd scenario (when we use two base stations) because the behavior remains the same for the other two scenarios. Here we want to show that, in spite of the poor QoS throughput, the EDCF takes a good way with the packet delays depending on the priority of the CBR flows. The worst case is shown in the flow3 and flow4 where the delays achieve very important values.

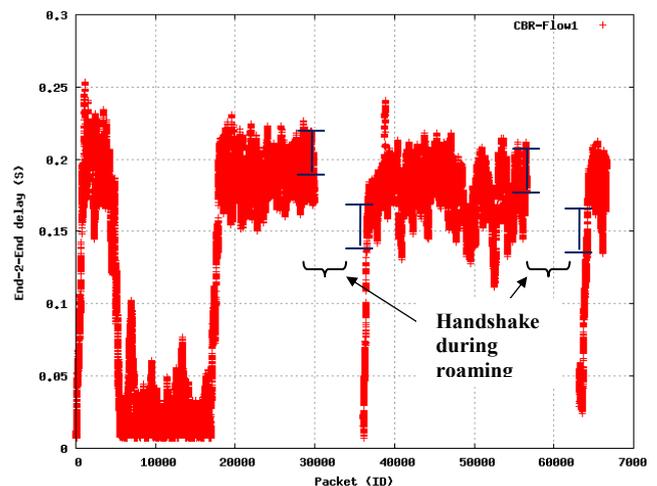

**Fig.2.A. End-2-end delay**

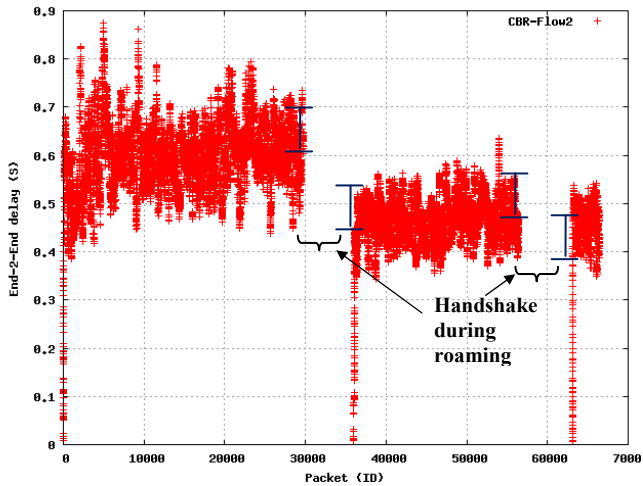

Fig.2.B. End-2-end delay

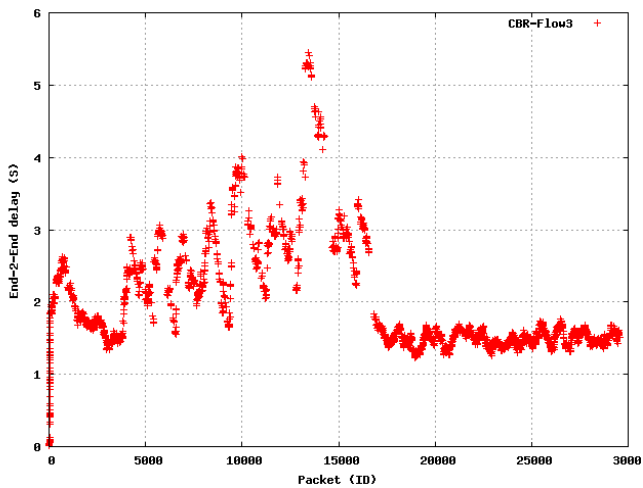

Fig.2.C. End-2-end delay

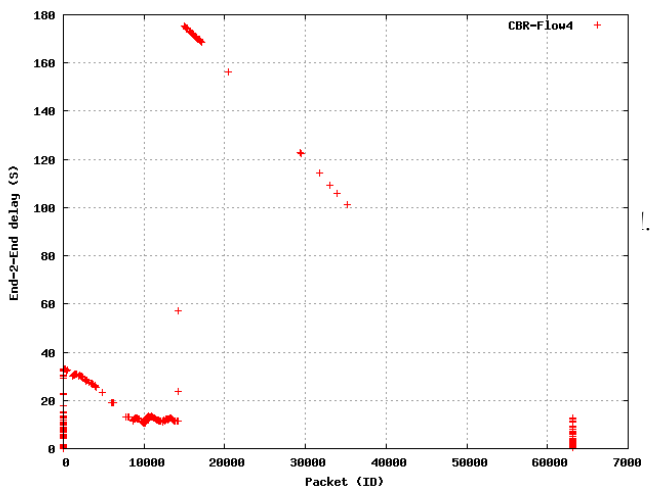

Fig.2.D. End-2-end delay

## IV. SCENARIOS-BASED ANALYSIS

Modeling on 1 Mbps, the wireless topology consisted of QSTA stations and a varying number of base stations connected as 5 Mbps gateways in the wired mode. In an IEEE 802.11e network in infrastructure mode, the mobile nodes always communicate directly with the base station.

The parameters for the wired link are chosen to ensure that the bandwidth bottleneck of the system is within the capacity of the wireless LAN. The mobile stations are located such that every base station is able to detect a transmission within its signal range, both in the case of mobility and in the case of staticity. Our simulations consist of traffic that has been chosen to be similar to those generated by a Constant Bit Rate (CBR) IP voice or H.264 as a real video encoder.

In the high priority traffic, the stations generate a flow which is taken from a normal distribution with mean 300 bytes, and standard deviation 40 bytes. We have used inter-packet intervals of 25 and 40 ms, which gives us data flows with an average bit rate of 60 kb/s. We will refer to these as high and low bit-rate.

In the low priority traffic, the stations generate packets every 50 ms, with a packet size taken from a normal distribution with mean 800 bytes, and standard deviation 150 bytes corresponding to a mean bit-rate of 128 kb/s.

We started measurements after a small period of time to allow the initial control traffic such as ARP to be exchanged, so, it would not affect the results of the simulation. We have tried to use the settings specified in the standards and the technical reports, where the schemes are precisely mentioned. Thus, we chose the settings and parameters to be used within different schemes. Table. 1 gives these values.

|  | Throughput (B/s) Static Scenario with 1 BS | Throughput (B/s) Mobile Scenario with 2 BS's | Number of Packets Received |
|---|---|---|---|
| Flow 1 | 60343 | 48610 | 52794 |
| Flow 2 | 19360 | 17682 | 19195 |
| Flow 3 | 6124 | 2988 | 2935 |
| Flow 4 | 580 | 247 | 249 |
| Total Packet Loss | <100 | 16880 |  |

Table.1. Parameter values for Fig.3

*Simulations*

Our simulation is designed around a hybrid structure of network (wireless and wired). Combining BS and MS, three scenarios are proposed:

*A. Base-Station and Static Wireless Nodes*

In the first scenario (Fig.3A), by allocating a fixed position to wireless node, the EDCF operates easily and accurately on the basis of service differentiation between the flows. The separation is very clear between the traffics throughput levels. This is mainly due to the static topology of the network which highly allows the scheduler to process correctly.

*B. Base-Station and Mobile Wireless Nodes*

In the second scenario (Fig. 3B), we activated the motion of the node with the speed of 20 m/s. We observe that EDCF looses immediately the ability of the traffic differentiation, and the flows seam scheduled as in the older DCF. Unlike the situation when the flows contending for medium access, the order of the occupied duration priority is still respected. This behavior is specially occurred by node's mobility phenomenon.

*C. Two Base-Stations and Mobile Wireless Nodes*

In the third scenario (Fig. 3C), which is bit complicated, the motion of the wireless node is traced by two Base Stations, which are spaced at 120°. Then increasing their range, quality detection is performed. Depending on the path, the mobile node traces, within one BS, the result remains the same as previously. The attractive phenomenon occurs when the mobile node switches to the second BS. If the ranges are not overlapped, the packets are automatically dropped. The traffic interruption, generated by the disconnection from the first BS and connection to the second BS and vice-versa (shown by the 2-way roaming handshake) which justifies the important number of packet loss in this mode (see Table. 1). We can observe again that when BS2 is switched on, the response rate is quite good and the throughput increases till its highest level (58000 bps) before disconnecting twice on the path of the back-way.

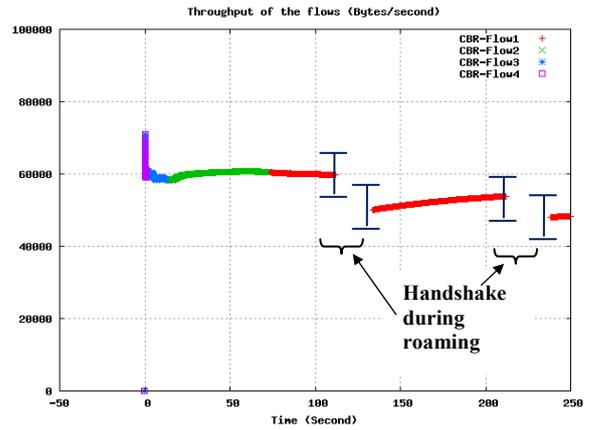

**Fig. 3.C. Throughput for scenario C**

## V. CONCLUSION AND FUTURE WORK

In this paper, we have performed a study of IEEE 802.11e QoS: EDCF-Contention-based in both static and dynamic Scenarios. We investigated the three MAC layer metrics; throughput, end-2-end delay and packet loss for three different scenarios. In static case, the EDCF operates with its full strength comparing to the dynamic case within the same topology, where the QoS scheme looses the differentiation service ability and comes down to the conventional functionality except ordering the access priority. By introducing roaming among BS's, the standard handoff's keep satisfying for throughput response. We plane to work on the overlaps of BS's with multi-ranges to investigate the mobility within the MAC protocol stability domain. We would like to extend the study over more scenarios following a couple of routing metrics which are already discussed and analyzed in previous work [6].

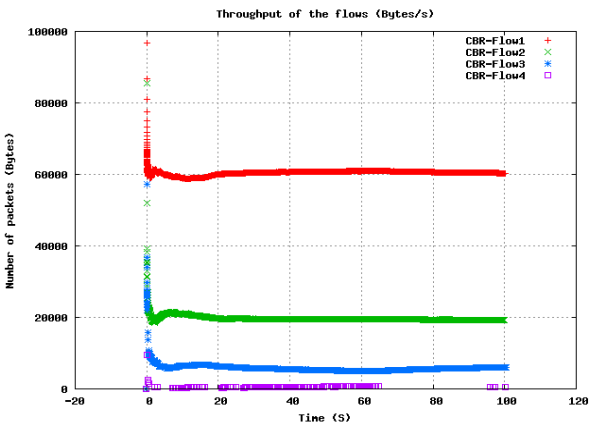

**Fig.3.A. Throughput for scenario A**

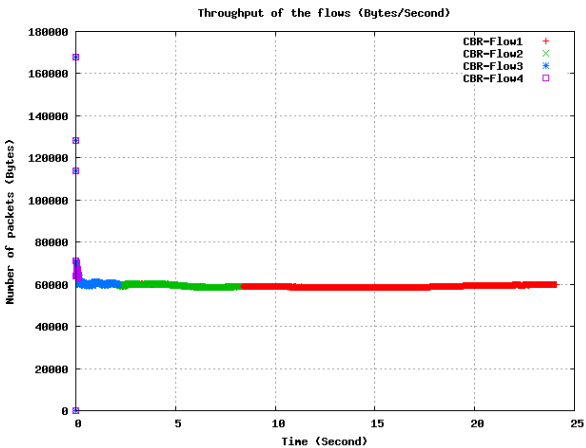

**Fig. 3.B. Throughput for scenario B**